\documentclass[twocolumn,pra, aps]{revtex4}

\usepackage{lineno}
\usepackage{subfigure}
\usepackage{sublabel}
\usepackage{dcolumn}
\usepackage{amsmath,amssymb}
\usepackage{bm}
\usepackage{bbm}
\usepackage{color}
\usepackage{overpic}
\usepackage{latexsym}
\usepackage{epstopdf}
\usepackage{color}
\usepackage[english]{babel}
\usepackage{times}
\usepackage{latexsym}
\usepackage{stmaryrd}
\usepackage{psfrag,graphicx}
\usepackage{epsf}
\usepackage{subfigure}
\usepackage{amsmath}
\usepackage{amssymb}
\usepackage{amsfonts}
\usepackage{natbib}
\usepackage{epstopdf}
\usepackage{appendix}
\usepackage{ulem}

\definecolor{mygrey}{gray}{0.35}
\definecolor{myblue}{rgb}{0.2,0.2,0.8}
\definecolor{myzard}{cmyk}{0,0,0.05,0}
\definecolor{mywhite}{rgb}{1,1,1}
\definecolor{myred}{rgb}{0.9,0.1,0.}
\usepackage[colorlinks=true,citecolor=myblue,linkcolor=myblue,urlcolor=myblue]{hyperref}

\def\be{\begin{equation}}
\def\ee{\end{equation}}
\def\ba{\begin{align}}
\def\enda{\end{align}}
\def\bi{\begin{itemize}}
\def\ei{\end{itemize}}

\newcommand{\tr}{\operatorname{tr}} 
\newcommand{\argmin}{\operatorname{argmin}} 
\newcommand{\id}{\mathbbm{1}}

\begin{document}


\title{A Scalable Maximum Likelihood Method for Quantum State Tomography}

\author{T.\ Baumgratz}
\author{A.\ N\"{u}\ss eler}
\author{M.\ Cramer}
\author{M.B.\ Plenio}

\affiliation{Institut f\"{u}r Theoretische Physik, Albert-Einstein-Allee 11,
Universit\"{a}t Ulm, 89069 Ulm, Germany}

\begin{abstract}
The principle of maximum likelihood reconstruction has proven to yield satisfactory results in the context of quantum state tomography for many-body systems of moderate system sizes. Until recently, however, quantum state tomography has been considered to be infeasible for systems consisting of a large number of subsystems due to the exponential growth of the Hilbert space dimension with the number of constituents. Several reconstruction schemes have been proposed since then to overcome the two main obstacles in quantum many-body tomography: experiment time and post-processing resources. Here we discuss one strategy to address these limitations for the maximum likelihood principle by adopting a particular state representation to merge a well established reconstruction algorithm maximizing the likelihood with techniques known from quantum many-body theory.
\end{abstract}

\maketitle

\section{Introduction}
The ability to prepare and manipulate quantum mechanical states is crucial for implementing quantum information processing and hence for building quantum computers \cite{divincenzo00}. Thus, it is important to verify the algorithms realized on quantum mechanical systems by conducting measurements on the system. This is achieved by means of quantum state tomography~\cite{hradil04,vogel89}, quantum process tomography \cite{chuang97} and quantum detector tomography \cite{lundeen09,zhang12} which together are capable of characterizing the three main stages of a quantum experiment. Here we concentrate on the task of quantum state tomography for an informationally incomplete set of observables. Deducing a quantum mechanical description of the system from these measurements is by no means trivial and what is the most appropriate method to invert the experimental data is still a matter of controversy \cite{rehacek07,blumekohout10}. The strategies of quantum state tomography range from directly inverting the measurements~\cite{james01} to statistical approaches known from classical probability theory such as maximum likelihood estimation \cite{james01,hradil97} and Bayesian inference \cite{blumekohout10,rau10}. The latter methods where introduced into the field of quantum state tomography to circumvent the possibility of negative eigenvalues of the state estimation arising by inverting Born's rule for relative frequencies, i.e., noisy data. In principle, maximum likelihood estimation and Bayesian inference divide the approach to statistical inference into two schools. The former is the method at hand for the frequentist statisticians where probabilities are interpreted as the infinite limit of relative frequencies obtained in a measurement process. Principally, the probabilities can be determined solely by repeating the measurements on the system infinitely often. For Bayesian statisticians, probabilities are a degree of belief in a certain event. The prior probability distribution imposed by the observer is then updated exploiting additional knowledge (i.e., measurements) of the system. Both principles have their advantages and drawbacks in the field of quantum state tomography. While maximum likelihood estimation is biased towards rank deficient states, no prior knowledge is required. In contrast, in Bayesian inference it is often not clear how to choose the prior distribution, whereas the estimates will have full rank \cite{blumekohout10,rau10}. In this manuscript, we will focus on the frequentist approach to quantum state tomography.

Maximum likelihood estimation is currently the method of choice and has proven to yield satisfactory results for moderate system sizes \cite{haeffner05}. However, as the system sizes realized in the laboratory steadily increase \cite{haeffner05,leibfried05,monz11,yao12,Islam13}, new techniques for performing quantum state tomography on large system sizes are required due to the exponential growth both in the number of required measurements and in the classical post-processing that is required to connect the measured data with a description of the quantum state. Recently, several scalable schemes for verifying the state in the laboratory by a fidelity estimation or learning the state via a concrete state reconstruction have been presented \cite{gross10,gross11,dasilva11,flammia11,flammia12,landoncardinal12,toth10,moroder12,cramer10,baumgratz12}. The efficiency of the scalable schemes stems from the exploitation of additional structure of the considered states such as a low rank, a local nature of correlations or a specific form of the state. If these assumptions are true, they allow for a reduction of the number of parameters which have to be determined experimentally and, simultaneously, for a reduction of the experiment time. To obtain an estimate of the state within these models, the scalable tomography schemes come along with efficiently implementable reconstruction algorithms to reduce the required post-processing resources. The algorithms are designed, for instance, to optimize a fit function over pure states~\cite{cramer10}, to adopt maximum likelihood estimation to permutationally invariant states, which can be represented efficiently \cite{moroder12}, or to invert local measurements to obtain a (not necessarily positive) state estimate \cite{baumgratz12}.

In this manuscript we strive to push the applicability of large scale state tomography further towards an efficient implementation of a maximum likelihood estimation scheme exploiting both the spirit of maximum likelihood reconstruction and adopting a specific representation of the states. We are considering one-dimensional quantum systems composed of $N$ finite-dimensional $d$-level subsystems (e.g., qubits with $d=2$). To realize the scalable tomography scheme we show how to implement a well established fixed point algorithm~\cite{rehacek07} maximizing the likelihood function by means of matrix product states and operators \cite{oestlund95,schollwoeck11,perezgarcia07}. Note that in this manuscript we restrict to the iterative fixed point algorithm for maximizing the likelihood function. One could, however, also adopt a direct maximization of the likelihood function using convex optimization tools together with a suitable representation of the considered states (this has been demonstrated, e.g., for permutationally invariant systems in \cite{moroder12}).
This manuscript is organized as follows. We start by reviewing the maximum likelihood approach to quantum state tomography in section~\ref{sec:QuantumStateTomographyviaMaximumLikelihoodEstimation}. Then, we introduce the fixed point algorithm and discuss a modification of the latter when a pure state estimate is desired. In section \ref{sec:MergingMatrixProductOperatorsWiththeMaximumLikelihoodAlgorithm} we show that all operations necessary for executing the maximum likelihood algorithm can be implemented by means of matrix product states and operators. Here, we solely discuss the individual computational steps and postpone details to the Appendix. If a certain operator comprising the POVM elements of the measurement setting has a matrix product operator representation of low bond-dimension, we show how
this can be done efficiently.
 In the Appendix we discuss how this operator can be recast to fit into the framework of matrix product states and operators for two specific measurement settings.
In the last section of this manuscript, section \ref{sec:NumericalExperiments}, we present numerical results demonstrating the performance of the algorithm for numerically simulated states.

\section{\label{sec:QuantumStateTomographyviaMaximumLikelihoodEstimation}Quantum State Tomography via Maximum Likelihood Estimation}

In this section we review the maximum likelihood approach to quantum state tomography and discuss a fixed point algorithm which maximizes the likelihood function \cite{hradil04,rehacek07}. We further motivate a modification of the algorithm when a pure state estimate is desired. The latter reformulation reduces the computational cost of the algorithm since it only requires matrix--vector multiplications instead of matrix--matrix multiplications.

Quantum state tomography is a procedure for estimating the density matrix from repeated measurements on $M$ identically prepared quantum systems. Let $\{\hat{\Pi}_{i}\}$, $i\in\Delta$, be the set of all POVM elements corresponding to the measurements performed on the system where $\Delta$ is an index set associated to these measurements \cite{remark1}. In the measurement process we record the number of times $n_{i}$ the outcome $\hat{\Pi}_{i}$ is obtained for all $i\in\Delta$, i.e., $M=\sum_{i\in\Delta} n_{i}$. Now, let $\mathcal{D}$  be the state space of the physical system under consideration. According to Born's rule, the conditional probability of measuring outcome $i$ given state $\hat{\varrho}\in\mathcal{D}$ is equal to $p_{i} = p(\hat{\Pi}_{i} \vert \hat{\varrho}) = \text{tr}[\hat{\Pi}_{i}\hat{\varrho}]$.
The joint probability $ p({\bf n} \vert \hat{\varrho})$  of registering the specific outcomes $n_{i}$, $i\in\Delta$, is given by
\begin{equation}
\mathcal{L}(\hat{\varrho}) = p({\bf n} \vert \hat{\varrho}) = \prod_{i\in\Delta} p_{i}^{n_{i}} = \prod_{i\in\Delta} \text{tr}[\hat{\Pi}_{i}\hat{\varrho}]^{n_{i}}.
\end{equation}
The likelihood function $\mathcal{L}(\hat{\varrho})$ is interpreted as the conditional probability of measuring on the state $\hat{\varrho}$ when registering outcome ${\bf n}$, i.e., it is considered to be a function on $\mathcal{D}$ for a specific outcome of a measurement series. The maximum likelihood estimate $\hat{\varrho}_{\text{ML}}$ is the element of the state space $\mathcal{D}$ which maximizes the likelihood function. Therefore, it is considered as the most likely state given a set of measurement outcomes.

Instead of maximizing the likelihood function $\mathcal{L}(\hat{\varrho})$ directly, one maximizes the log-likelihood function
\begin{equation}
\text{log} \, \mathcal{L}(\hat{\varrho}) = \sum_{i\in\Delta} n_{i} \, \text{log}(\text{tr}[\hat{\Pi}_{i}\hat{\varrho}]),
\end{equation}
which yields the same estimate $\hat{\varrho}_{\text{ML}}$ due to the fact that the likelihood function $\mathcal{L}(\hat{\varrho})$ is positive on the state space and the logarithm is strictly increasing. The resulting function reveals the property of being concave and since we are optimizing over a convex set we are left with a convex optimization problem \cite{boyd09}.
It is well known that the state $\hat{\varrho}_{\text{ML}}$ maximizing the log-likelihood satisfies $\hat{\varrho}_{\text{ML}} = \hat{R}(\hat{\varrho}_{ML})\hat{\varrho}_{\text{ML}}$ where the positive operator $\hat{R}(\hat{\varrho})$ is given by \cite{hradil04}
\begin{equation}
\hat{R}(\hat{\varrho}) = \frac{1}{M} \sum_{i\in\Delta} \frac{n_{i}}{p_{i}} \hat{\Pi}_{i}  = \frac{1}{M} \sum_{i\in\Delta} \frac{n_{i}}{ \text{tr}[\hat{\Pi}_{i}\hat{\varrho}] } \hat{\Pi}_{i}.
\label{eqn:Roperator}
\end{equation}
Since both the operator $\hat{R}(\hat{\varrho})$ and $\hat{\varrho}$ are Hermitian, the extremal equation $\hat{\varrho}_{\text{ML}} = \hat{R}(\hat{\varrho}_{ML})\hat{\varrho}_{\text{ML}}$ can be transformed to yield a well-established fixed point algorithm~\cite{hradil04}
\begin{equation}
\hat{\varrho}_{k+1} = \mathcal{N}\left[ \hat{R}(\hat{\varrho}_{k}) \hat{\varrho}_{k} \hat{R}(\hat{\varrho}_{k}) \right]
\label{eqn:MLfixedpointalgorithm}
\end{equation}
where the operation $\hat{R}(\hat{\varrho}_{k}) \hat{\varrho}_{k} \hat{R}(\hat{\varrho}_{k})$ preserves the positivity of the current iteration, the function $\mathcal{N}:\mathcal{H}^{d^N \times d^N}\rightarrow \mathcal{D}$ mapping Hermitian matrices to states denotes normalization to trace one and the algorithm is usually initialized by an unbiased initial state such as the completely mixed state. Note that, although heuristically convergent, there is no analytical proof that the fixed point algorithm \eqref{eqn:MLfixedpointalgorithm} converges to the maximum of the log-likelihood function. Diluting the operator $\hat{R}(\hat{\varrho})$ to
$(\id + \epsilon \hat{R}(\hat{\varrho}) )/ (1+\epsilon)$, one can show that for $\epsilon \ll 1$ the log-likelihood increases in each step of the fixed point algorithm at the expense of the rate of convergence \cite{rehacek07}. Since the log-likelihood function is concave, this guarantees that the diluted fixed point algorithm converges to the maximum.

Under the assumption that the state in the laboratory is pure, the fixed point algorithm can be rewritten to optimize only over pure states. In particular, iterating
\begin{equation}
|\psi_{k+1}\rangle = \mathcal{N}\left[ \hat{R}(|\psi_{k}\rangle) |\psi_{k}\rangle \right]
\label{eqn:MLfixedpointalgorithmpurestates}
\end{equation}
will yield a pure estimate of the desired state. Note, however, that restricting the state space to rank one matrices alters the optimization problem to being non-convex. This implies that although the objective function, i.e., the log-likelihood, is still concave the feasible set, i.e., the set of rank one matrices, is non-convex and hence it is not guaranteed that the global maximum is attained by increasing the log-likelihood in every step, both in the standard formulation and the diluted version of this pure state algorithm. Nevertheless, as we will see, iterating only over pure states yields satisfactory results in practice even for noisy data. Now, of course, in a realistic setting the state prepared in the laboratory will never be pure. A pure state estimate, however, can nevertheless be used to certify whether the (possibly mixed) state implemented in the laboratory is close to this pure estimate by means of the techniques described in~\cite{cramer10}. With this, one can construct a lower bound on the fidelity of the pure state estimate with respect to the actual state only by means of the (experimentally determined) reduced density matrices.

\section{\label{sec:MergingMatrixProductOperatorsWiththeMaximumLikelihoodAlgorithm}Merging Matrix Product Operators With the Maximum Likelihood Algorithm}

In this section, we introduce the family of matrix product states and operators \cite{oestlund95,schollwoeck11,perezgarcia07} and further show that the iterative fixed point algorithm \eqref{eqn:MLfixedpointalgorithm} can be implemented using this formalism. We only give a brief summary of the individual steps and postpone details to the Appendix.

Let $\hat{\varrho}\in\mathcal{D}$ be a state describing a physical system comprising $N$ $d$-level subsystems. Then, $\hat{\varrho}$ can be written as
\begin{equation}
\hat{\varrho}=\sum_{\alpha_1,\dots,\alpha_N}P_1[\alpha_1]\cdots P_N[\alpha_N] \;\hat{P}_1^{(\alpha_1)}\cdots \hat{P}_N^{(\alpha_N)},
\label{eqn:mporepresentation}
\end{equation}
where $\{\hat{P}_{k}^{(\alpha_{k})}\}$ is an operator basis of $\mathbb{C}^{d\times d}$ for site $k$, $\alpha_{k} = 1,\ldots,d^2$ enumerates the basis elements per site and $P_{k}[\alpha_{k}] \in\mathbb{C}^{D_{k} \times D_{k+1}}$ are complex matrices for $k=1,\ldots,N$ with $D_{1} = 1$ and $D_{N+1}=1$. Every state can be brought to this form and the exponentially growing dimension of the Hilbert space (with respect to the number of subsystems $N$) can be covered by exponentially growing matrix dimensions $D_{k}$. However, many interesting quantum states (in particular living on one-dimensional structures) can be represented with a moderate number of parameters.
Examples, amongst others, are ground states and thermal states of gapped local Hamiltonians \cite{hastings07,plenio05,eisert10,hastings06,audenaert02,brandao12}. These states have in common that they can be approximated very well by states with low bond-dimension $D = \max_{k} D_{k}$. We call states of the form of equation \eqref{eqn:mporepresentation} matrix product operators or matrix product states depending on $\{\hat{P}_{k}^{(\alpha_{k})}\}$ being a basis of $\mathbb{C}^{d\times d}$ or $\mathbb{C}^{d}$ \cite{schollwoeck11,perezgarcia07} (note that every matrix product state with bond-dimension $D$ is a matrix product operator with bond-dimension at most $D^2$) and collect such states with maximum bond-dimension equal to $D$ in the set $\mathcal{M}_D$.

Next, we list the operations required to run the algorithm and then proceed by pointing out the corresponding matrix product operator equivalents. In what follows, we restrict ourselves to a POVM which contains solely observables of tensor product form, i.e., $\hat{\Pi}_{i} =\hat{\pi}_1^{i}\otimes\cdots\otimes \hat{\pi}_N^{i}$ with $i\in\Delta$. The fixed point algorithm is commonly initialized with the completely mixed state. Then, in the $k^{\text{th}}$ iteration, the algorithm:
\begin{enumerate}
\item[1.] determines the expectation values of the POVM elements $\{\hat{\Pi}_{i}\}$, $i\in\Delta$, i.e., $p_{i} = \tr[\hat{\Pi}_{i}\hat{\varrho}_{k}]$,
\item[2.] constructs the operator $\hat{R}(\hat{\varrho}_{k})$,
\item[3.] multiplies the current iteration $\hat{\varrho}_{k}$ from both sides with $\hat{R}(\hat{\varrho}_{k})$, and
\item[4.] normalizes the output to obtain iteration $k+1$.
\end{enumerate}

Whenever $\hat{\varrho}_k$, $\hat{R}(\hat{\varrho}_{k})$ and hence $\hat{R}(\hat{\varrho}_{k})\hat{\varrho}_k\hat{R}(\hat{\varrho}_{k})$ may be written as matrix product operators of small bond-dimension, all these steps may be performed efficiently (for a detailed outline of the different operations that are required we refer to the literature \cite{schollwoeck11,perezgarcia07} and Appendix~\ref{sec:Appendix1}):

We now outline how the four steps above may be performed efficiently.
First, computing expectation values of product observables of the form $\hat{\Pi} = \hat{\pi}_{1}\otimes\cdots\otimes \hat{\pi}_{N}$ is straightforward:
\begin{equation}
\tr[\hat{\varrho} \,\hat{\Pi}] = \prod_{k=1}^{N} E_{k}
\label{eqn:MPOexpectationvalue}
\end{equation}
where $E_{k} = \sum_{\alpha_{k}=1}^{d^2} P_{k}[\alpha_{k}] \; \tr[\hat{\pi}_{k}\hat{P}_{k}^{(\alpha_{k})}]$.

Secondly, the operator $\hat{R}(\hat{\varrho}_{k})$ has to be determined. This operator acts on the state space and hence, generally, has the same dimension as $\hat{\varrho}_{k}$. To efficiently implement the maximum likelihood algorithm we therefore need to represent this operator as a matrix product operator with small bond-dimension. Since $\hat{R}(\hat{\varrho}_{k})$ is basically a weighted sum of the POVM elements, this restricts the set of measurements for which this is possible. To give an example for which
this can be achieved, consider POVM elements $\hat{\Pi}_k^{i}$ of the form
\begin{equation}
\hat{\Pi}^{i}_{k}=\id\otimes\cdots\otimes\id\otimes\hat{\pi}_k^{i}\otimes\cdots\otimes\hat{\pi}_{k+R-1}^{i}\otimes\id\otimes\cdots\otimes\id
\end{equation}
where $k=1,\ldots,N-R+1$ labels all blocks of $R$ consecutive spins and $i$ enumerates the set of local POVM elements. Key point for the efficient representation of the operator $\hat{R}(\hat{\varrho}_{k})$ is that with this specific kind of measurement set-up this operator is in fact a local Hamiltonian. But local Hamiltonians are matrix product operators of low bond-dimension and hence $\hat{R}(\hat{\varrho}_{k})$ can be stored efficiently in its matrix product operator representation (see Appendix \ref{sec:Appendix2} for further details). Note again that every measurement setting where the operator $\hat{R}(\hat{\varrho}_{k})$ has an efficient matrix product operator representation with small bond-dimension is suitable for the efficient implementation of the maximum likelihood algorithm and the above form is merely an example. One may also add global observables to the POVM set, a scenario which we consider for the reconstruction of a GHZ-type state below, with a discussion of the construction of the operator $\hat{R}(\hat{\varrho}_{k})$ in Appendix \ref{sec:Appendix3}.

Thirdly, to compute the next iteration of the algorithm~\eqref{eqn:MLfixedpointalgorithm} we need to multiply the current iteration $\hat{\varrho}_{k}$ with the operator $\hat{R}(\hat{\varrho}_{k})$ from both sides. Both objects are stored as matrix product operators. Multiplying two matrix product operators with bond-dimensions $D_{1}$ and $D_{2}$ results in a matrix product operator of dimension $D_{1} \cdot D_{2}$, i.e., bond-dimensions multiply. To keep the bond-dimension at a certain level (e.g., $D$), we compress the current iteration $\hat{R}(\hat{\varrho}_{k})\hat{\varrho}_{k}\hat{R}(\hat{\varrho}_{k})$ that has been obtained after multiplication of the previous estimate $\hat{\varrho}_{k}$ with the operator $\hat{R}(\hat{\varrho}_{k})$, i.e., we solve \cite{schollwoeck11}
\begin{equation}
\hat{\xi}_{k} = \argmin[ \|  \hat{\varrho} -  \hat{R}(\hat{\varrho}_{k}) \hat{\varrho}_{k} \hat{R}(\hat{\varrho}_{k})\|^2 \; \vert \; \hat{\varrho}\in\mathcal{M}_{D}],
\label{eqn:MPOcompression}
\end{equation}
where $\|\cdot \|$ denotes the Hilbert-Schmidt norm and $\mathcal{M}_{D}$ comprises all matrix product operators with bond-dimension $D$, see Appendix \ref{sec:Appendix1} for a detailed discussion. Obviously, solving equation \eqref{eqn:MPOcompression} is a crucial point in the algorithm. If the operator is not compressible to a low bond-dimension, the algorithm will not be efficient. To recognize this issue in the execution of the algorithm, we choose a method which allows us to track the error made in the compression, i.e., the algorithm aborts when the minimum in equation \eqref{eqn:MPOcompression} is not zero (or greater than a prescribed tolerance). Hence, if the norm difference does not converge to zero, we either choose a larger bond-dimension or abort the maximum likelihood algorithm (i.e., throw a flag and abort the tomography scheme).

Fourthly, in the final step of one iteration we have to normalize the current estimate to trace one. Again, this can be done efficiently since the trace of a matrix product operator can be computed directly using equation \eqref{eqn:MPOexpectationvalue}. Normalization is then equivalent to dividing each matrix by a fraction of the initial trace.

\section{\label{sec:NumericalExperiments}Numerical Simulations}

In this section we present results of our implementation of the maximum likelihood algorithm iterating over matrix product states and operators~\cite{remark2}. The results suggest that applying maximum likelihood reconstruction to system sizes where the conventional implementations of full tomography fail due to~(i) the number of required measurement settings and (ii) the limited post-processing resources is still manageable for appropriate POVM settings.

\begin{figure}[tb]
	\begin{center}
		\includegraphics[width=0.9\columnwidth]{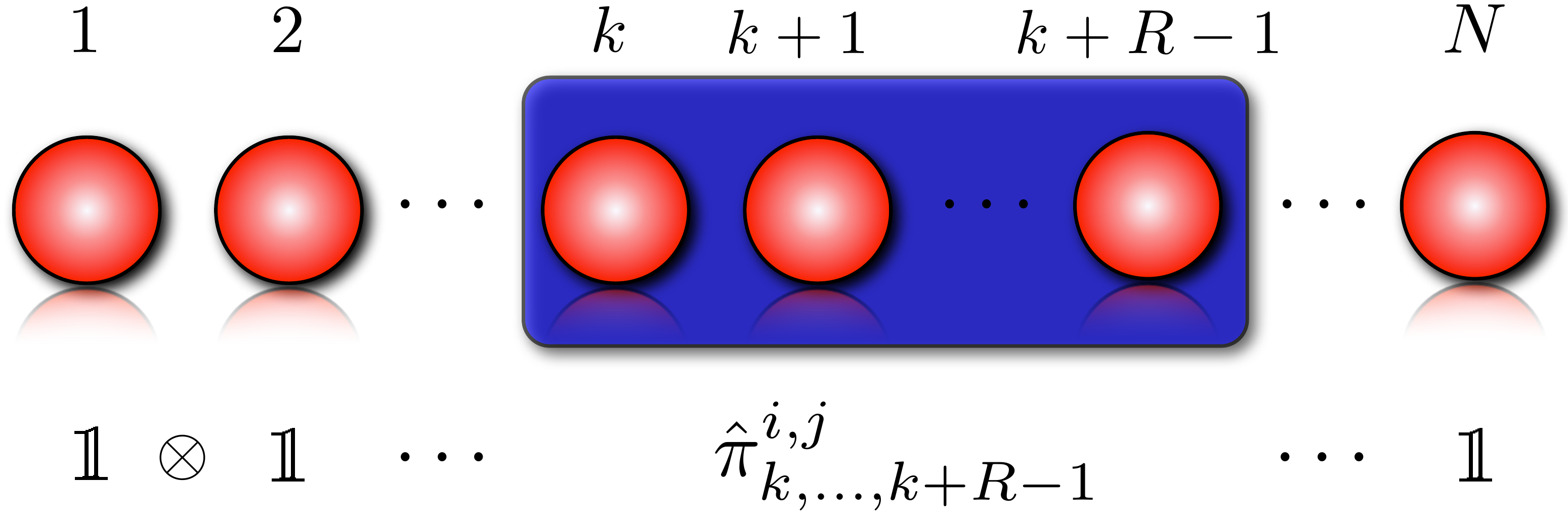}
	\end{center}
	\caption{The considered measurements on blocks of $R$ consecutive spins. For $\hat{\pi}^{i,j}_{k,\ldots,k+R-1}$ see  equation~(\ref{eqn:localPOVMelement}).
	Here, $k=1,\ldots,N-R+1$ labels the block of size $R$, $i=1,\ldots,3^R$ enumerates the basis rotation (all combinations of orientations along the $X,Y$ and $Z$ direction) and $j=1,\ldots,2^R$ denotes the corresponding projectors per basis orientation on the eigenbasis of the respective Pauli spin operators.}
\label{fig:MeasurementOperators}
\end{figure}

We are considering a one-dimensional chain of quantum systems implemented on $N$ $d$-level systems with $d=2$, i.e., qubits. We let the POVM elements be of the local form
\begin{equation}
\hat{\Pi}^{i,j}_{k} = \mathcal{W}(\id_{1,\ldots,k-1} \otimes \hat{\pi}_{k,\ldots,k+R-1}^{i,j} \otimes \id_{k+R,\ldots,N}),
\end{equation}
see figure~\ref{fig:MeasurementOperators}. Here,
\begin{equation}
\hat{\pi}^{i,j}_{k,\ldots,k+R-1} = |y^{i,j}_{k,\ldots,k+R-1} \rangle \langle y^{i,j}_{k,\ldots,k+R-1}|,
\label{eqn:localPOVMelement}
\end{equation}
with $k=1,\ldots,N-R+1$ labelling the block of size $R$, $i=1,\ldots,3^R$ enumerating the basis rotation (all combinations of orientations along the $X,Y$ and $Z$ direction) and $j=1,\ldots,2^R$ denoting the corresponding projectors per basis orientation on the eigenbasis of the respective Pauli spin operators. This corresponds to measurements
on blocks of $R$ contiguous spins. For fixed $R$, the experimental effort, and hence the associated experiment time, grows linearly in the number of subsystems $N$ reducing the exponential scaling of the total number of measurements from $M = 3^{N}m$ to a linearly scaling $M=3^{R}m(N-R+1)$. Here, $m$ denotes the number of measurements per basis orientation.

We verify the performance of the maximum likelihood algorithm optimizing over matrix product operators (and matrix product states) for thermal states and ground states of next-neighbour Hamiltonians and for GHZ-type states. For the latter, we need to enlarge the above POVM, resulting in a set of measurements determining the GHZ-type state uniquely and a total number of measurements $M=3^{R}m(N-R+1) + Km$, where $K$ is the cardinality of the set of additional (global) measurements.

We simulate the measurements by drawing $m$ times from the exact multinomial distributions corresponding to the POVM elements specifying each basis orientation. Hence, the input to the reconstruction scheme are the relative frequencies $f^{i,j}_{k} = n^{i,j}_{k} / m$ for all $i,j$ and $k$ approximating the exact probability distributions. The reconstructed states $\hat{\varrho}_{\text{rec}}$ are compared to the exact states by means of the renormalized Hilbert-Schmidt norm difference $D(\hat{\varrho},\hat{\varrho}_{\text{rec}}) = \|\hat{\varrho} - \hat{\varrho}_{\text{rec}} \|^2 / \| \hat{\varrho}\|^2$, the fidelity $f(|\psi\rangle,\hat{\varrho}_{\text{rec}}) = | \langle\psi|\hat{\varrho}_{\text{rec}}|\psi\rangle |$ or $f(|\psi\rangle,|\psi_{\text{rec}}\rangle) = |\langle\psi|\psi_{\text{rec}}\rangle|^2$.

\begin{figure}[tb]
	\begin{center}
		\includegraphics[width=0.9\columnwidth]{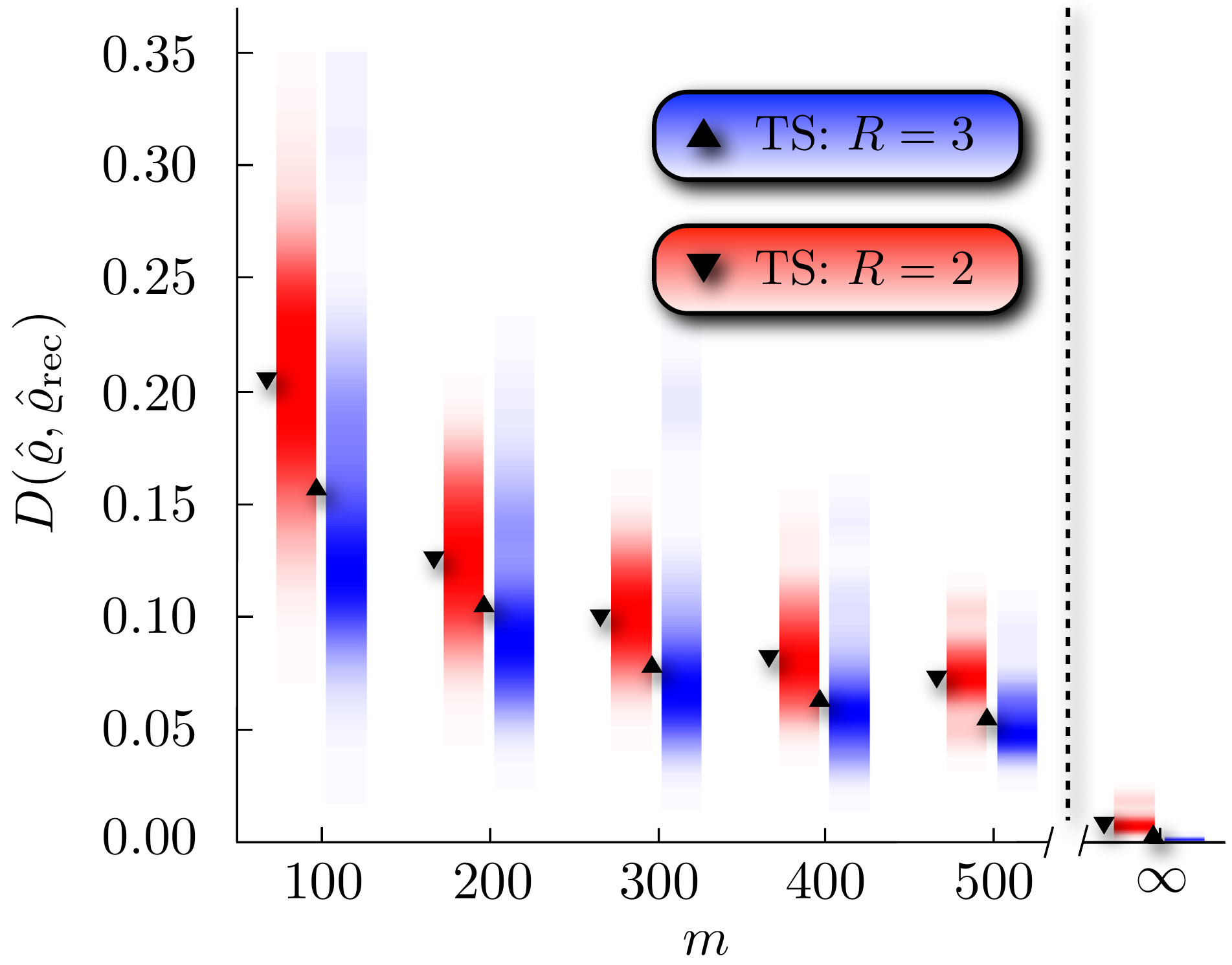}
	\end{center}
	\caption{Performance of the iterative algorithm \eqref{eqn:MLfixedpointalgorithm} for thermal states~(TS) of random next-neighbour Hamiltonians of the form of equation~\eqref{eqn:randomNNHamiltonian} for $N=16$ qubits with $\beta = 2$ after 1000 iterations. Input to the reconstruction scheme are relative frequencies corresponding to a tomographically complete set of measurements on all blocks of $R$ contiguous sites. These relative frequencies are obtained by drawing from the exact probability distributions $m$ times. Downward-pointing triangles: $R=2$, upward-pointing triangles: $R=3$. The densities illustrate the results for $45$ random Hamiltonians and corresponding reconstructions while the triangles emphasize the mean value. All estimates have a bond-dimension of $D=16$.}
\label{fig:RandomNNThermalStates}
\end{figure}

The first example outlines the performance of the algorithm~\eqref{eqn:MLfixedpointalgorithm} iterating over matrix product operators for thermal states of random next-neighbour Hamiltonians,

\begin{equation}
\hat{H} = \sum_{i=1}^{N-1} \hat{r}_{i,i+1},
\label{eqn:randomNNHamiltonian}
\end{equation}
where the Hermitian matrices $\hat{r}_{i,i+1}$ act on sites $i$ and $i+1$. The real and imaginary parts of these matrices are drawn from a Gaussian distribution with zero mean and standard deviation one. Thermal states $\hat{\varrho}=e^{- \beta  \hat{H}}/Z$ are obtained by an imaginary time evolution using the time evolving block-decimation algorithm \cite{zwolak04,remark3}. From these states, we compute the exact probability distributions corresponding to all contiguous blocks of size $R$ for local tomographically complete measurements, simulate the measurements and reconstruct a state estimate $\hat{\varrho}_{\text{rec}}$. Results are shown in figure~\ref{fig:RandomNNThermalStates}. The plot indicates that, as expected,~(i) the accuracy of the reconstruction scheme increases with an increasing number of measurements and~(ii)~extending the range $R$ on which measurements are performed decreases the error of the reconstructed estimate. Notably, the mean error for the reconstruction of thermal states corresponding to $45$ different random Hamiltonians of the form of equation~\eqref{eqn:randomNNHamiltonian} with $R=3$ and perfect measurements, i.e., $m=\infty$, is of order $10^{-3}$.

\begin{figure}[htb]
	\begin{center}
		\includegraphics[width=0.9\columnwidth]{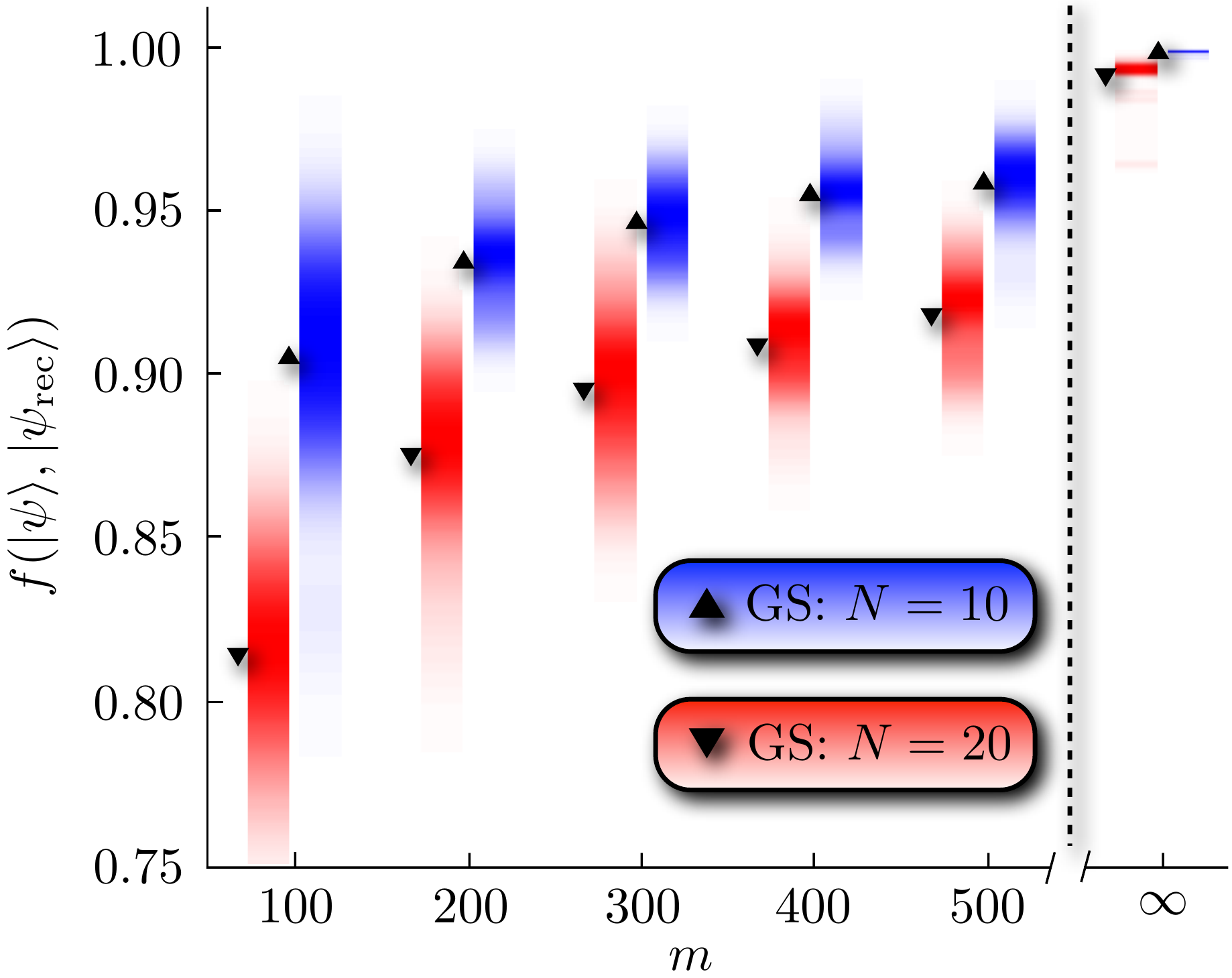}
	\end{center}
	\caption{Results for the maximum likelihood algorithm iterating over matrix product states after $5000$ iterations. The exact states are ground states (GS) of random next-neighbour Hamiltonians of the form of equation~\eqref{eqn:randomNNHamiltonian}. Note that the reconstructed estimates have a bond-dimension of $D=5$. Downward-pointing triangles: $N=20$ sites, upward-pointing triangles: $N=10$ sites. Measurements are simulated for all blocks of $R=2$ contiguous sites by drawing from the exact multinomial distributions $m$ times. For each $m$ we generate $30$ random Hamiltonians. The densities depict the distributions of the obtained fidelities while triangles indicate the mean value. Here, for $N=20$, one iteration of the algorithm takes about one second on a standard laptop. }
\label{fig:RandomNNGroundStates}
\end{figure}

Ground states of Hamiltonians of the form of equation~\eqref{eqn:randomNNHamiltonian} serve as our second example. We obtain the ground states by minimizing the expectation value of the Hamiltonian with respect to a matrix product state with given bond-dimension~\cite{remark4}. Sweeping through the chain and optimizing the matrices in the matrix product state site by site such that in each iteration the expectation value with respect to the Hamiltonian decreases, we will end up in a stationary point which serves as an approximation to the exact ground state \cite{schollwoeck11,remark5}. Again, we compute the exact expectation values corresponding to local measurements on all blocks of $R$ sites, simulate the measurements by drawing $m$ times from the exact multinomial distributions and reconstruct the state using the maximum likelihood algorithm. Here, we are optimizing only over pure states using the iterative algorithm in equation~\eqref{eqn:MLfixedpointalgorithmpurestates}. Figure~\ref{fig:RandomNNGroundStates} presents the results. Even for very small $m$ we obtain a mean fidelity larger than $0.80$ for $N=20$ qubits. Moreover, for the exact probabilities ($m=\infty$) the mean fidelity is close to $1.00$ after $5000$ iterations indicating that iterating only over pure states does not get stuck in local minima. Note that the iterative algorithm is known to converge very slowly in comparison to other maximization techniques for the likelihood function \cite{moroder12}.

\begin{figure}[htb]
	\begin{center}
		\includegraphics[width=0.9\columnwidth]{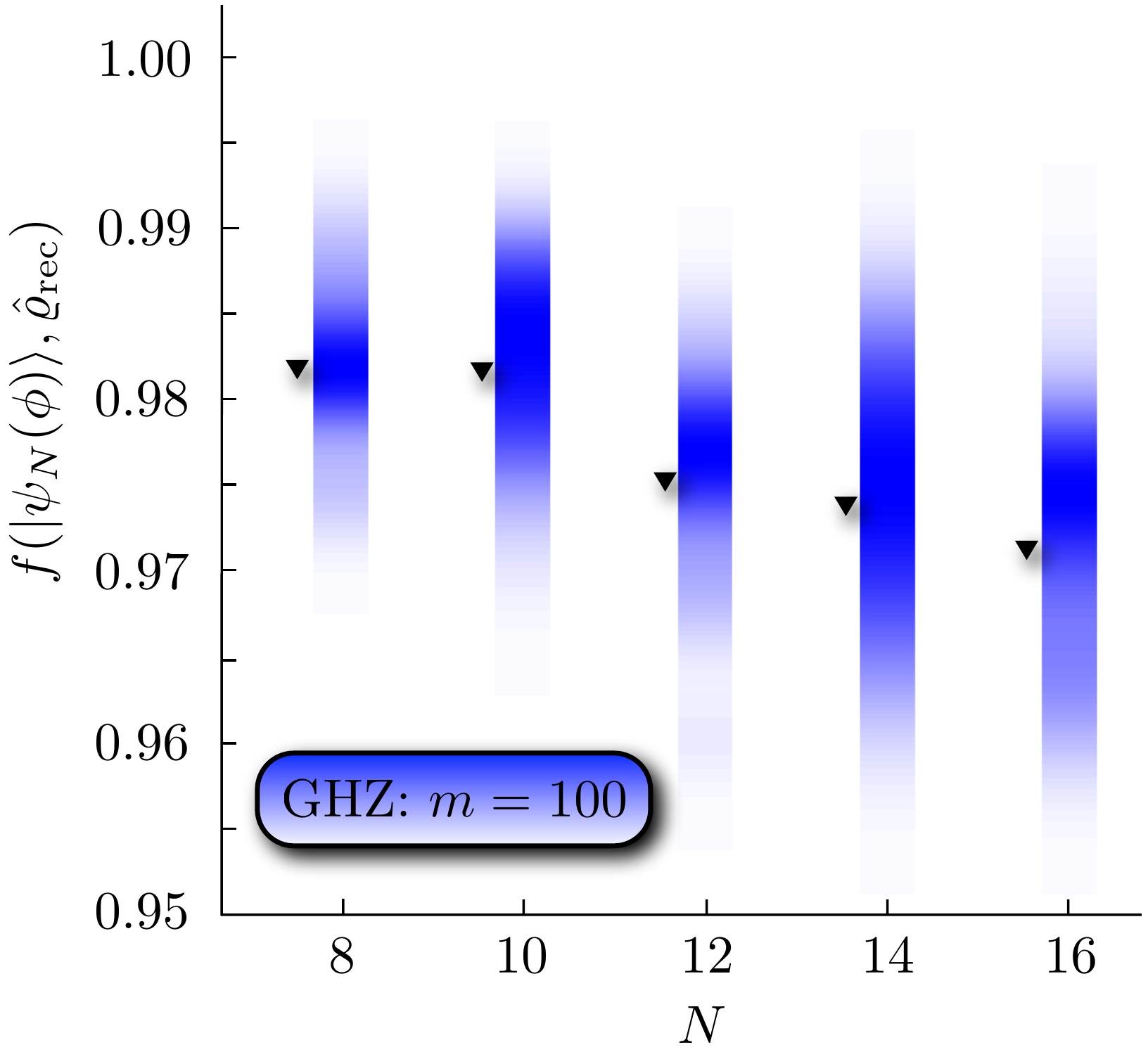}
	\end{center}
	\caption{Results for GHZ-type states as shown in equation~\eqref{eqn:GHZtypeStates} where we choose $\phi = \pi/2$. Apart from the local measurements (on all blocks of size $R=2$) global observables have to be measured given by $\{\hat{\Pi}_{i}\}$, $i\in\Delta_{2}$. Here, we draw $m=100$ times from the exact probability distributions and reconstruct 30 states per $N$. Mean values are indicated by the downward-pointing triangles. The reconstructed estimates have bond-dimension $D=10$. The plot shows the fidelity after $1000$ iterations of the maximum likelihood algorithm. }
\label{fig:GHZtypeState}
\end{figure}

Finally, let us consider GHZ-type states of the form
\begin{equation}
|\psi_{N}(\phi)\rangle = [|0\rangle^{\otimes N/2} |1\rangle^{\otimes N/2} + e^{i\phi} |1\rangle^{\otimes N/2}\rangle |0\rangle^{\otimes N/2}] / \sqrt{2}
\label{eqn:GHZtypeStates}
\end{equation}
where the number of sites $N$ is even and $\phi \in [0;2\pi)$ is a relative phase. These states are not uniquely determined by local measurements. The relative phase $\phi$ is a global feature and hence, besides local measurements, we need to perform some global measurements to fully determine the state. This could be done by additionally measuring the operators $\hat{X}^{\otimes N}$ and $\hat{Y}\otimes \hat{X}^{\otimes N-1}$ since the expectation values
\begin{equation}
\langle \psi_{N}(\phi) |\hat{X}^{\otimes N}| \psi_{N}(\phi) \rangle = \cos(\phi)
\label{eqn:GHZPOVM1}
\end{equation}
and
\begin{equation}
\langle \psi_{N}(\phi) |\hat{Y}\otimes \hat{X}^{\otimes N-1}| \psi_{N}(\phi) \rangle = \sin(\phi)
\label{eqn:GHZPOVM2}
\end{equation}
fix the local phase. Note that these expectation values can be determined experimentally for large system sizes because only a simultaneous rotation of all spins is required plus a single site addressing to rotate one spin (e.g., the spin on the very left-hand side) along an orthogonal direction. We define the POVM $\{\hat{\Pi}_{i}\}$, $i=1,\ldots,4$, with $\hat{\Pi}_{1} = (\id + \hat{X}^{\otimes N})/4$, $\hat{\Pi}_{2} = (\id - \hat{X}^{\otimes N})/4$, $\hat{\Pi}_{3} = (\id + \hat{Y} \otimes \hat{X}^{\otimes N-1})/4$ and $\hat{\Pi}_{4} = (\id - \hat{Y} \otimes \hat{X}^{\otimes N-1})/4$. These operators are positive and sum to one. Incorporating these operators into the POVM describing the local measurements can be done by a straightforward normalization of the operators. Let us denote the index set of the full POVM set by $\Delta = \Delta_{1} \cup \Delta_{2}$ where $\Delta_{1}$ comprises the elements corresponding to local measurements and $\Delta_{2}$ comprises the four elements defined above. To apply the maximum likelihood algorithm for matrix product operators to this measurement scheme, it remains to show that the operator $\hat{R}(\hat{\varrho})$, see equation~\eqref{eqn:Roperator}, can be written as a matrix product operator with small bond-dimension $D$. In fact, we find
\begin{equation}
\hat{R}(\hat{\varrho}) = \hat{R}_{1}(\hat{\varrho}) + \hat{R}_{2}(\hat{\varrho})
\end{equation}
where $\hat{R}_{k}(\hat{\varrho}) = \frac{1}{M} \sum_{i\in\Delta_{k}} \frac{n_{i}}{p_{i}} \hat{\Pi}_{i}$ for $k=1,2$. As before, $\hat{R}_{1}(\hat{\varrho})$ corresponds to a matrix product operator with small bond-dimension $D_{1}$ due to its local character. Moreover, $\hat{R}_{2}(\hat{\varrho})$ can be written as a matrix product operator with bond-dimension $D_{2} = 2$ such that $\hat{R}(\hat{\varrho})$ is a matrix product operator with bond-dimension $D = D_{1} + 2$ (see Appendix \ref{sec:Appendix3} for further details). We simulate measurements on the exact GHZ-type state (which is a matrix product state with bond-dimension $D_{\text{GHZ}} = 2$) for $R=2$ and use the algorithm iterating over matrix product operators to obtain a state estimate. Results are shown in figure \ref{fig:GHZtypeState}. This example illustrates that this method is also applicable to states that are not uniquely determined by local measurements only. In general, however, the experimentalist has to have some prior knowledge about the state he intends to implement on a physical system to decide whether additional measurements are necessary. The results suggest that even for a very small number of measurements where each considered basis rotation is measured $m=100$ times, the algorithm is capable of reconstructing GHZ-type states very accurately. 

\section{Conclusions}
We presented a scalable maximum likelihood algorithm for quantum state tomography. The only restrictions for the applicability is that the set of measurement operators is polynomial in the system size and that the individual POVM elements are of tensor product form, both of which are anyway generally desirable from an experimental point of view. The reconstruction technique relies on a well-established fixed point algorithm for maximizing the likelihood function. We have shown that this algorithm can be generalized to iterate over matrix product states and operators and hence yields a scalable reconstruction algorithm for quantum state tomography. Of course, for a general state where the measurements do not uniquely specify the state, the fixed point of the scalable maximum likelihood algorithm will not always come close to the true state. If the state is uniquely determined by the measured POVM elements, however, we provided numerical evidence that the algorithm chooses a state estimate which is close to the true state.

We observed that the convergence of the algorithm is very slow caused by the flat log-likelihood function when measurements are done only on a small subset of the full Hilbert space. Moreover, while our first and by no means optimal implementation is capable of dealing with measurements on blocks of size $R\geq2$ the execution of the algorithm becomes rapidly slow with increasing $R$. An improvement of the implementation is left for future work. For instance, one could think of directly maximizing the concave objective function (e.g., by gradient descend methods) when the states are represented as matrix product states or in any other tailored efficient form. Concretely, restricting to permutationally invariant states and directly maximizing the likelihood, \cite{moroder12} has shown to yield superior performance when high accuracy is required.

The numerical results, however, suggest that the generalization of the iterative algorithm to matrix product operators is able to find state estimates which are close to the exact states already for a small number of measurements, which we verified for thermal and ground states of random next-neighbour Hamiltonians and for GHZ-type states.
 Furthermore, restricting the algorithm to optimize only over pure states has shown to yield satisfactory results under the assumption that the desired state is pure. This is remarkable since one iterates over a non-convex set. To certify the reconstructed pure estimate one could use the techniques described in \cite{cramer10} to find a lower bound on the fidelity of the state in the laboratory with respect to the pure state estimate.
The results presented in this manuscript add into recent research for large scale quantum tomography and push the applicability of quantum state tomography to large system sizes.

We gratefully acknowledge R. Rosenbach for the supply of results of the TEBD algorithm. Computations were performed on the bwGRiD \cite{BWGrid}. This work was supported by the Alexander von Humboldt Foundation, the EU Integrating project SIQS, the EU STReP EQUAM and the BMBF Verbundprojekt QuOReP.



\onecolumngrid

\setcounter{equation}{0}
\renewcommand{\theequation}{A\arabic{equation}}
\numberwithin{equation}{section}

\appendix
\section{\label{sec:Appendix1}Implementation Details for the Maximum Likelihood Algorithm Iterating over Matrix Product Operators}
In the first part of the Appendix we review some basic properties and specific operations applied to matrix product operators (for a general overview we refer to \cite{schollwoeck11,perezgarcia07} where the ideas discussed in the remainder of this section of the Appendix are presented in far more detail). We focus on the operations required for generalizing the maximum likelihood fixed point algorithm.

To set the notation, we recall the definition of a matrix product operator. For all sites $k=1,\ldots,N$ and $\alpha_{k}=1,\ldots,d^2$, let $\{P_{k}[\alpha_{k}]\}$ be a set of complex matrices of dimension $D_{k} \times D_{k+1}$ where $D_{1} = 1 = D_{N+1}$. Further, let $\{\hat{P}_k^{(\alpha_k)}\}$ denote an operator basis for site $k$, e.g., for spin-$1/2$ particles the orthonormal Pauli spin basis, i.e., $\hat{P}^{(1)}_{i} = \id_{i} / \sqrt{2}$, $\hat{P}^{(2)}_{i} = \hat{X}_{i} / \sqrt{2}$, $\hat{P}^{(3)}_{i} = \hat{Y}_{i} / \sqrt{2}$ and $\hat{P}^{(4)}_{i} = \hat{Z}_{i} / \sqrt{2}$.
The state
\begin{equation}
\hat{\varrho}=\sum_{\alpha_1,\dots,\alpha_N}P_1[\alpha_1]\cdots P_N[\alpha_N] \; \hat{P}_1^{(\alpha_1)}\cdots \hat{P}_N^{(\alpha_N)}
\label{eqn:mporepresentation_appendix}
\end{equation}
is called matrix product operator if $D = \max_{k} D_{k}$ is low with respect to the state space dimension. Note, of course, that the notation of a low bond-dimension is a bit vague. The efficient simulation of a matrix product operator of bond-dimension $D$ certainly depends on the computational resources available for the post-processing.

\subsubsection*{Basic Matrix Product Operator Manipulations}
In the first subsection of Appendix \ref{sec:Appendix1} we present a derivation of equation~\eqref{eqn:MPOexpectationvalue} in the main text, yielding an efficient strategy for computing expectation values of product observables with respect to matrix product operators. Further, we show how to multiply matrix product operators and give a specific representation of the completely mixed state. Let $\hat{\Pi} = \hat{\pi}_{1}\otimes\cdots\otimes\hat{\pi}_{N}$ be a product observable acting on $N$ $d$-level subsystems. The expectation value of $\hat{\Pi}$ with respect to the matrix product operator $\hat{\varrho}$ is determined via
\begin{equation}
\begin{split}
\tr[\hat{\varrho}\hat{\Pi}] &= \sum_{\alpha_1,\ldots,\alpha_N} P_1[\alpha_1]\cdots P_N[\alpha_N] \; \tr[ \hat{P}_1^{(\alpha_1)}\cdots \hat{P}_N^{(\alpha_N)} \hat{\pi}_{1}\cdots\hat{\pi}_{N} ] \\
& = \sum_{\alpha_1,\ldots,\alpha_N} P_1[\alpha_1]\cdots P_N[\alpha_N]  \;  \tr[ (\hat{P}_1^{(\alpha_1)}\hat{\pi}_{1}) \cdots (\hat{P}_N^{(\alpha_N)} \hat{\pi}_{N})] \\
& = \sum_{\alpha_1,\ldots,\alpha_N} P_1[\alpha_1]\cdots P_N[\alpha_N]  \; \tr[ \hat{P}_1^{(\alpha_1)} \hat{\pi}_{1}] \cdots \tr[\hat{P}_N^{(\alpha_N)} \hat{\pi}_{N}] \\
&= \prod_{k=1}^{N} E_{k}
\end{split}
\end{equation}
where $E_{k} = \sum_{\alpha_{k}=1}^{d^2} P_{k}[\alpha_{k}] \; \tr[\hat{P}_{k}^{(\alpha_{k})}\hat{\pi}_{k}] \in \mathbb{C}^{D_{k} \times D_{k+1}}$. In the derivation we used the fact that for systems $i$ and $i+1$ we have  $\hat{A}_{i}\hat{B}_{i+1} \hat{C}_{i}\hat{D}_{i+1}= \hat{A}_{i}\otimes\hat{B}_{i+1} \cdot \hat{C}_{i} \otimes \hat{D}_{i+1} = \hat{A}_{i}\cdot\hat{C}_{i} \otimes\hat{B}_{i+1}\cdot\hat{D}_{i+1} = \hat{A}_{i}\hat{C}_{i}\hat{B}_{i+1}\hat{D}_{i+1}$ and that $\tr[\hat{A}_{i}\hat{B}_{i+1}] = \tr[\hat{A}_{i}\otimes \hat{B}_{i+1}] = \tr[\hat{A}_{i}]\tr[\hat{B}_{i+1}]$.

Naively multiplying two matrix product operators with bond-dimensions $D_{1}$ and $D_{2}$ yields a matrix product operator with bond-dimension at most $D_{1}\cdot D_{2}$. In particular, given the matrix product operators
\begin{equation}
\hat{\varrho}=\sum_{\alpha_1,\dots,\alpha_N}P_1[\alpha_1]\cdots P_N[\alpha_N] \; \hat{P}_1^{(\alpha_1)}\cdots \hat{P}_N^{(\alpha_N)}
\end{equation}
with bond-dimension $D_{1}$ and
\begin{equation}
\hat{\xi}=\sum_{\beta_1,\dots,\beta_N}Q_1[\beta_1]\cdots Q_N[\beta_N] \; \hat{P}_1^{(\beta_1)}\cdots \hat{P}_N^{(\beta_N)}
\end{equation}
with bond-dimension $D_{2}$, we find
\begin{equation}
\begin{split}
\hat{\varrho} \cdot \hat{\xi} &=\sum_{\alpha_1,\dots,\alpha_N} \sum_{\beta_1,\dots,\beta_N}P_1[\alpha_1]\cdots P_N[\alpha_N] \; Q_1[\beta_1]\cdots Q_N[\beta_N] \; \hat{P}_1^{(\alpha_1)}\cdots \hat{P}_N^{(\alpha_N)} \; \hat{P}_1^{(\beta_1)}\cdots \hat{P}_N^{(\beta_N)} \\
 &=\sum_{\alpha_1,\dots,\alpha_N} \sum_{\beta_1,\dots,\beta_N} (P_1[\alpha_1] \otimes Q_1[\beta_1]) \cdots (P_N[\alpha_N]\otimes Q_N[\alpha_N]) (\hat{P}_1^{(\alpha_1)} \hat{P}_1^{(\beta_1)})\cdots (\hat{P}_N^{(\alpha_N)} \hat{P}_N^{(\beta_N)} ). \\
\end{split}
\end{equation}
Now, for every site $k=1,\ldots,N$, $\{\hat{P}^{(\alpha_{k})}_{k}\}$ denotes an operator basis such that we have $\hat{P}_k^{(\alpha_k)} \hat{P}_k^{(\beta_k)} = \sum_{\gamma_k} c^{\alpha_k,\beta_k}_{\gamma_k} \hat{P}_{k}^{(\gamma_k)}$ with $c^{\alpha_k,\beta_k}_{\gamma_k} = \tr[(\hat{P}^{(\gamma_k)}_k)^\dagger \hat{P}^{(\alpha_k)}_k  \hat{P}^{(\beta_k)}_k ]$. Consequently, the product of two matrix product operators gives
\begin{equation}
\begin{split}
\hat{\varrho} \cdot \hat{\xi} &= \sum_{\gamma_1,\ldots,\gamma_N} \left(\sum_{\alpha_1,\beta_1} c^{\alpha_1,\beta_1}_{\gamma_1} P_1[\alpha_1] \otimes Q_1[\beta_1]\right) \cdots \left(\sum_{\alpha_N,\beta_N} c^{\alpha_N,\beta_N}_{\gamma_N} P_N[\alpha_N] \otimes Q_N[\beta_N]\right) \cdot \hat{P}_1^{(\gamma_1)}\cdots \hat{P}_N^{(\gamma_N)} \\
& = \sum_{\gamma_1,\ldots,\gamma_N} R_1[\gamma_1]\cdots R_N[\gamma_N] \; \hat{P}_1^{(\gamma_1)}\cdots \hat{P}_N^{(\gamma_N)}
\end{split}
\end{equation}
with $R_k[\gamma_k] = \sum_{\alpha_k,\beta_k}  \tr[(\hat{P}^{(\gamma_k)}_k)^\dagger \hat{P}^{(\alpha_k)}_k  \hat{P}^{(\beta_k)}_k ] \; P_k[\alpha_k] \otimes Q_k[\beta_k]$. Hence, the product of two matrix product operators with bond-dimensions $D_{1}$ and $D_{2}$ is itself a matrix product operator with bond-dimension at most $D = D_{1}\cdot D_{2}$.

The maximum likelihood algorithm is initialized by the completely mixed state. Choosing the orthonormal Pauli spin basis in the matrix product operator representation, this state corresponds to a matrix product operator with bond-dimension one. One possible choice of the $1\times 1$ matrices defining the completely mixed state is
\begin{equation}
\begin{array}{lllll}
P_{k}[1] 		& = & \frac{1}{N\cdot \sqrt{d^N}}	&,	&  \\
P_{k}[\alpha_k]	& = & 0 \hspace{0.5cm}		&, \text{ for } \alpha_k=2,\ldots,d^2 \text{ and }k=1,\ldots,N.
\end{array}
\end{equation}
Therefore, initializing the matrix product operator formulation of the maximum likelihood algorithm is straightforward.

\subsubsection*{Compressing  Matrix Product Operators}

We have seen that the bond-dimension of the current iteration $\hat{\varrho}_{k}$ increases in each step of the algorithm due to the matrix product operator multiplication with the operator $\hat{R}(\hat{\varrho}_{k})$. Hence, it is essential to introduce subroutines to keep the bond-dimension of the state estimate low. This is done by compressing the matrix product state to a state with smaller bond-dimension at each iteration of the tomography algorithm \cite{schollwoeck11}. In the second subsection of Appendix \ref{sec:Appendix1} we describe the compression of a matrix product operator $\hat{\varrho}$ with bond-dimension $D_{1}$ to a matrix product operator $\hat{\sigma}$ with bond-dimension $D_{2} < D_{1}$. For that, we show how to solve the optimization problem
\begin{equation}
\hat{\sigma} = \argmin[ \|  \hat{\varrho} -  \hat{\varrho}^{\prime}\|^2 \; \vert \; \hat{\varrho}^{\prime}\in\mathcal{M}_{D_2}],
\end{equation}
see equation~\eqref{eqn:MPOcompression} in the main text.

Before we describe the iterative compression scheme, let us first introduce a convenient representation of the matrix product operator. For this, note that the matrices $\{ P_{k}[\alpha_{k}] \}$ are not unique since inserting $\id = UU^{\dagger}$ between sites $k$ and $k+1$ will not alter the state, but the corresponding matrices $\{ P_{k}[\alpha_{k}] \}$ and $\{ P_{k+1}[\alpha_{k+1}] \}$. This freedom in the choice of the matrices allows (by a successive singular value decomposition starting from the left- and right-hand side of the chain) to write the matrix product operator as
\begin{equation}
\hat{\varrho} = \sum_{\alpha_1,\ldots,\alpha_N} B_{1}[\alpha_1]\cdots B_{N}[\alpha_N] \; \hat{P}^{(\alpha_1)}_{1}\cdots \hat{P}^{(\alpha_N)}_{N}
\label{eqn:knormalMPO}
\end{equation}
where the matrices on sites $l = 1,\ldots,k-1$ satisfy
\begin{equation}
\sum_{\alpha_l} B_{l}[\alpha_l]^{\dagger}B_{l}[\alpha_l] = \id_{D_{l+1} \times D_{l+1}}
\end{equation}
and where the matrices on sites $l = k+1,\ldots,N$ satisfy
\begin{equation}
\sum_{\alpha_l} B_{l}[\alpha_l]B_{l}[\alpha_l]^{\dagger} = \id_{D_{l} \times D_{l}}.
\end{equation}
We call matrix product operators satisfying these relations k-normal matrix product operators \cite{schollwoeck11}. Further, note that for k-normal matrix product operators the purity, i.e., the squared Hilbert-Schmidt norm $\| \hat{\varrho} \|^2 = \tr[\hat{\varrho}^{\dagger}\hat{\varrho}]$, can be computed by
\begin{equation}
\| \hat{\varrho} \|^2 = \sum_{\alpha_k} \tr[B_{k}[\alpha_k]^{\dagger} B_{k}[\alpha_k]].
\end{equation}
Now, let
\begin{equation}
\hat{\varrho} = \sum_{\beta_{1},\ldots,\beta_{N}} A_{1}[\beta_1]\cdots A_{N}[\beta_N] \; \hat{P}^{(\beta_1)}_{1}\cdots \hat{P}^{(\beta_N)}_{N}
\end{equation}
be a matrix product operator with bond-dimension $D_{1}$ (not necessarily in k-normal form) and $\hat{\sigma}$ be a matrix product operator with bond-dimension $D_{2} < D_{1}$ satisfying
\begin{equation}
\hat{\sigma} = \argmin[ \|  \hat{ \varrho} - \hat{\varrho}^{\prime} \|^{2} \; \vert \; \hat{\varrho}^{\prime} \in \mathcal{M}_{D_{2}} ],
\end{equation}
i.e., $\hat{\sigma}$ is the matrix product operator with bond-dimension $D_{2} < D_{1}$ closest to $\hat{\varrho}$ with respect to the Hilbert-Schmidt norm (hence a compression of the latter). To determine $\hat{\sigma}$, we use a procedure where the minimization is performed iteratively by sweeping through the chain several times while optimizing the compressed matrices site by site \cite{schollwoeck11}, i.e., we minimize
\begin{equation}
\|  \hat{\varrho} - \hat{\varrho}^{\prime} \|^{2} = \tr[\hat{\varrho}^{\dagger}\hat{\varrho}] - \tr[\hat{\varrho}^{\dagger}\hat{\varrho}^{\prime} ] - \tr[(\hat{\varrho}^{\prime})^{\dagger}\hat{\varrho} ]  + \tr[(\hat{\varrho}^{\prime})^{\dagger}\hat{\varrho}^{\prime} ]
\label{eqn:normdistancetwoMPO}
\end{equation}
with respect to the complex conjugate matrices $\{B_{k}[\alpha_k]^{*}\}$  on site $k$ of the state
\begin{equation}
\hat{\varrho}^{\prime} = \sum_{\alpha_{1},\ldots,\alpha_{N}} B_{1}[\alpha_1]\cdots B_{N}[\alpha_N] \; \hat{P}^{(\alpha_1)}_{1}\cdots \hat{P}^{(\alpha_N)}_{N}
\end{equation}
given in k-normal form while keeping all the other matrices fixed. Note that only the last two terms in \eqref{eqn:normdistancetwoMPO} contribute to the minimization such that
\begin{equation}
\frac{\partial}{\partial (B_{k}[\alpha_{k}]^{*})_{i_{k},i_{k+1}}} \;\; \|  \hat{\varrho} - \hat{\varrho}^{\prime}  \|^{2} = \frac{\partial}{\partial (B_{k}[\alpha_{k}]^{*})_{i_{k},i_{k+1}}} \;\; ( - \tr[(\hat{\varrho}^{\prime})^{\dagger}\hat{\varrho} ]  + \tr[(\hat{\varrho}^{\prime})^{\dagger}\hat{\varrho}^{\prime} ]).
\end{equation}
Since the state $\hat{\varrho}^{\prime}$ is given in k-normal form, we find
\begin{equation}
\frac{\partial}{\partial (B_{k}[\alpha_{k}]^{*})_{i_{k},i_{k+1}}} \;\; ( \tr[(\hat{\varrho}^{\prime})^{\dagger}\hat{\varrho}^{\prime} ] ) = (B_{k}[\alpha_{k}])_{i_k,i_{k+1}}
\end{equation}
and
\begin{equation}
\frac{\partial}{\partial (B_{k}[\alpha_{k}]^{*})_{i_{k},i_{k+1}}} \;\; ( - \tr[(\hat{\varrho}^{\prime})^{\dagger}\hat{\varrho} ] ) = -(L_{1,\ldots,k-1}\cdot A_{k}[\alpha_{k}] \cdot R_{k+1,\ldots,N})_{i_{k},i_{k+1}}
\end{equation}
where
\begin{equation}
L_{1,\ldots,k-1} = \left( \sum_{\alpha_{k-1}} (B_{k-1}[\alpha_{k-1}])^{\dagger} \left( \ldots \left( \sum_{\alpha_{1}} (B_{1}[\alpha_{1}])^{\dagger}  A_{1}[\alpha_{1}]\right) \ldots \right) A_{k-1}[\alpha_{k-1}] \right)
\end{equation}
and
\begin{equation}
R_{k+1,\ldots,N} = \left( \sum_{\alpha_{k+1}} A_{k+1}[\alpha_{k+1}] \left( \ldots \left( \sum_{\alpha_{N}} A_{N}[\alpha_{N}] (B_{N}[\alpha_{N}])^{\dagger} \right) \ldots \right)  (B_{k+1}[\alpha_{k+1}])^{\dagger} \right).
\end{equation}
Hence, the extremal point is given by
\begin{equation}
B_{k}[\alpha_{k}] = L_{1,\ldots,k-1}\cdot A_{k}[\alpha_{k}] \cdot R_{k+1,\ldots,N}.
\label{eqn:updaterule}
\end{equation}
Finally, the iterative minimization of the norm difference can be performed by starting with a N-normal matrix product operator of dimension $D_{2}$, updating the matrices $L_{1,\ldots,N-1}$ and $\{B_{N}[\alpha_{N}]\}$ and a successive (N$-1$)-normalization of the resulting matrix product operator. The matrices on site $N-1, \ldots, 1$ are subsequently updated by the rule~\eqref{eqn:updaterule} while one needs to keep the matrix product operator in its k-normal form when optimizing on site $k$. Sweeping through the chain back and forth several times will lead to a convergence of this procedure. Since we are minimizing the norm distance in an iterative manner, this scheme might get stuck in local minima \cite{schollwoeck11}. Therefore we monitor the norm difference by exploiting that after updating the matrices on site $k$ we have
\begin{equation}
\|  \hat{\varrho} - \hat{\varrho}^{\prime}  \|^{2} = \| \hat{\varrho} \|^{2} - \sum_{\alpha_{k}} \tr[B_{k}[\alpha_{k}](B_{k}[\alpha_{k}])^{\dagger}].
\end{equation}
With this, we can either abort the algorithm if the norm difference does not converge to zero or increase the bond-dimension and redo the compression. To avoid the attraction of local minima, one can consider two sites instead of only one site in each step of the compression algorithm. Decomposing the optimized matrix products living on two sites by means of a singular value decomposition yields the matrices living on the single sites and hence the matrix product structure is preserved \cite{schollwoeck11}.

\section{\label{sec:Appendix2}R-local Hamiltonians are Matrix Product Operators}

In this section of the Appendix we are considering the specific measurement setting where all POVM elements act non-trivially only on subsets of $R$ consecutive sites. In this scenario, the operator $\hat{R}(\hat{\varrho})$ has the structure of a R-local Hamiltonian and hence a matrix product operator representation where the bond-dimension can be related to the interaction range $R$.

It is crucial for the scalability of the matrix product operator formulation of the maximum likelihood algorithm, see equation~\eqref{eqn:MLfixedpointalgorithm} of the main text, that the operator \cite{hradil04}
\begin{equation}
\hat{R}(\hat{\varrho}) = \frac{1}{M} \sum_{i\in\Delta_{1}} \frac{n_{i}}{p_{i}} \hat{\Pi}_{i}  = \frac{1}{M} \sum_{i\in\Delta_{1}} \frac{n_{i}}{ \text{tr}[\hat{\Pi}_{i}\hat{\varrho}] } \hat{\Pi}_{i}
\end{equation}
can be written as a matrix product operator of low bond-dimension. Now, let the POVM elements $\{\hat{\Pi}_{i}\}$, $i\in\Delta_{1}$, only act non-trivially on a subset of $R$ consecutive sites. Then, we find
\begin{equation}
\hat{R}(\hat{\varrho}) = \sum_{i=1}^{N-R+1} \sum_{\alpha_{1},\ldots,\alpha_{R}=1}^{n}  c_{i}[\alpha_{1},\ldots,\alpha_{R}] \;\hat{S}^{(\alpha_{1})}_{i} \ldots \hat{S}^{(\alpha_{R})}_{i+R-1}
\label{eqn:klocalRrepresentation}
\end{equation}
where we consider $n$ operators on each site and comprise all coefficients of the operator $\hat{R}(\hat{\varrho})$ in $c_{i}[\alpha_{1},\ldots,\alpha_{R}]$. Note, that the operator $\hat{S}^{(\alpha_{1})}_{i} \cdots \hat{S}^{(\alpha_{R})}_{i+R-1}$ only acts on sites $i,i+1,\ldots,i+R-1$ and that, for the moment, we do not require that $\{\hat{S}^{(\alpha_{k})}_{k}\}$ is a basis on site $k=1,\ldots,N$. In this measurement setting, the operator $\hat{R}(\hat{\varrho})$ obeys the form of a R-local Hamiltonian. It is well known that R-local Hamiltonians are matrix product operators \cite{schollwoeck11}. Before we show how to convert a R-local Hamiltonian into its matrix product operator representation, let us discuss an alternative representation of matrix product operators. Let $\hat{\varrho}$ be a matrix product operator of bond-dimension $D$. Then \cite{schollwoeck11}
\begin{equation}
\begin{split}
\hat{\varrho} &=\sum_{\alpha_1,\dots,\alpha_N}P_1[\alpha_1]\cdots P_N[\alpha_N] \; \hat{P}_1^{(\alpha_1)}\cdots \hat{P}_N^{(\alpha_N)} \\
&= \sum_{\alpha_1,\dots,\alpha_N} \sum_{i_{2},\ldots,i_{N}} (P_{1}[\alpha_{1}])_{1,i_{2}} (P_{2}[\alpha_{2}])_{i_{2},i_{3}} \cdots (P_{N}[\alpha_{N}])_{i_{N},1} \; \hat{P}_1^{(\alpha_1)}\cdots \hat{P}_N^{(\alpha_N)} \\
& = \sum_{i_{2},\ldots,i_{N}} \left( \sum_{\alpha_{1}} (P_{1}[\alpha_{1}])_{1,i_{2}} \hat{P}_{1}^{(\alpha_{1})} \right)\cdots \left( \sum_{\alpha_{N}} (P_{N}[\alpha_{N}])_{i_{N},1} \hat{P}_{1}^{(\alpha_{N})} \right) \\
& = \hat{\varrho}^{[1]}\ldots\hat{\varrho}^{[N]}
\end{split}
\end{equation}
where the operator-valued matrix $(\hat{\varrho}^{[k]})_{i_{k},i_{k+1}} = \left( \sum_{\alpha_{k}} (P_{k}[\alpha_{k}])_{i_{k},i_{k+1}} \hat{P}_{k}^{(\alpha_{k})} \right)$ acts only on system $k$ and where $i_{1}=1$ and $i_{N+1}=1$. We will refer to this representation of the matrix product operators as the operator-valued matrix product operator representation. Note that, given the operator-valued matrices, one can find the matrices of the matrix product operator representation by inversion, i.e.,
\begin{equation}
(P_{k}[\alpha_{k}])_{i_{k},i_{k+1}} = \tr[\hat{P}_{k}^{(\alpha_{k})}  (\hat{\varrho}^{[k]})_{i_{k},i_{k+1}}].
\label{eqn:ConvertOperatorValuedMPOinStandardMPO}
\end{equation}
We now set out to find an operator-valued matrix product operator representation of the R-local Hamiltonian $\hat{R}(\hat{\varrho})$, see equation~\eqref{eqn:klocalRrepresentation}. For this, we rewrite equation~\eqref{eqn:klocalRrepresentation} as
\begin{equation}
\begin{split}
\hat{R}(\hat{\varrho}) &= \sum_{i=1}^{N-R+1} \sum_{\alpha_{1},\ldots,\alpha_{R}=1}^{n}  c_{i}[\alpha_{1},\ldots,\alpha_{R}] \; \hat{S}^{(\alpha_{1})}_{i} \hat{S}^{(\alpha_{2})}_{i+1} \ldots \hat{S}^{(\alpha_{R})}_{i+R-1} \\
&=\sum_{i=1}^{N-R+1} \sum_{\alpha_{2},\ldots,\alpha_{R}=1}^{n}   \hat{Q}^{(\alpha_{2},\ldots,\alpha_{R})}_{i} \hat{S}^{(\alpha_{2})}_{i+1} \ldots \hat{S}^{(\alpha_{R})}_{i+R-1}
\end{split}
\end{equation}
where $\hat{Q}^{(\alpha_{2},\ldots,\alpha_{R})}_{i}  = \sum_{\alpha_{1}=1}^{n} c_{i}[\alpha_{1},\ldots,\alpha_{R}] \;\hat{S}^{(\alpha_{1})}_{i}$ for all $i=1,\ldots,N-R+1$. Let us illustrate the operator-valued matrices for $\hat{R}(\hat{\varrho})$ in the case of next-neighbour interaction, i.e., we measure only on all blocks of two contiguous sites and hence $R=2$. It is easily verified that the operator-valued matrices take on the form \cite{schollwoeck11}
\begin{equation}
\hat{R}^{[1]} =
\begin{bmatrix}
0			& \hat{Q}^{(1)}_{1}	& \hat{Q}^{(2)}_{1} 	& 	\ldots 	& \hat{Q}^{(n)}_{1}	& \id 	\\
\end{bmatrix}
\in\mathbb{C}^{d \times d(n+2) }
\end{equation}
for site $1$,
\begin{equation}
\hat{R}^{[k]} =
\begin{bmatrix}
\id 			& 0 				& 				&  	\ldots 	&  				& 0		\\
\hat{S}^{(1)}	& 0 				& 				&  	\ldots 	& 				& 0		\\
\hat{S}^{(2)}	& \vdots  			& 				&		 	& 				& \vdots	\\
\vdots 		& 	 			& 				&  	\ddots	& 				& 		\\
	 		& 	 			& 				&  			& 				& 		\\
\hat{S}^{(n)}	& 0 		 		& 				&  	\ldots	& 				& 0		\\
0			& \hat{Q}^{(1)}_{k}	& \hat{Q}^{(2)}_{k} 	& 	\ldots 	& \hat{Q}^{(n)}_{k}	& \id 	\\
\end{bmatrix}
\in\mathbb{C}^{d(n+2) \times d(n+2)}
\end{equation}
for sites $2,\ldots,N-1$ and
\begin{equation}
\hat{R}^{[N]} =
\begin{bmatrix}
\id 			\\
\hat{S}^{(1)}	\\
\hat{S}^{(2)}	\\
\vdots 		\\
	 		\\
\hat{S}^{(n)}	\\
0			\\
\end{bmatrix}
\in\mathbb{C}^{d(n+2) \times d}
\end{equation}
for site $N$, such that
\begin{equation}
\hat{R}(\hat{\varrho}) = \hat{R}^{[1]}\cdots\hat{R}^{[N]}.
\end{equation}
 The rules to obtain the operator-valued matrices for R-local Hamiltonians can be generalized straightforwardly. The dimensions of the resulting operator-matrices grow according to $d(2 + \sum_{i=1}^{R-1}n^{i})$ where $d$ is the on-site dimension, $n$ the number of considered operators per site (i.e., the cardinality of the set $\{\hat{S}^{(\alpha)}\}$) and $R$ the number of consecutive sites in one block on which measurements are performed. Equation~\eqref{eqn:ConvertOperatorValuedMPOinStandardMPO} allows to compute the matrices of the standard matrix product operator representation resulting in a bond-dimension of at most $d(2 + \sum_{i=1}^{R-1}n^{i})$. For the quantum state tomography of qubits ($d=2$) we have $n=6$ since every basis rotation (orientation along X,Y and Z) allows for two spin orientations (spin up or down in the respective basis). Consequently, for realistic quantum state tomography settings the bond-dimension of the operator $\hat{R}(\hat{\varrho})$ still grows rapidly when measured on large block sizes $R$. Heuristically, it turns out that the so constructed matrix product operator representation of the operator $\hat{R}(\hat{\varrho})$ is not optimal and that one can find a representation with smaller bond-dimension by compressing the corresponding matrices.

Note that with the strategy discussed above one constructs an exact matrix product operator representation of the operator $\hat{R}(\hat{\varrho})$. Of course, to obtain a matrix product operator approximation one could simply add the individual terms of the operator $\hat{R}(\hat{\varrho})$ which are often of matrix product operator structure or straightforwardly converted into the latter. Here, the bond-dimensions of the individual terms add to the overall bond-dimension. To keep the bond-dimension at a certain level, one could compress this operator as described in Appendix \ref{sec:Appendix1} to keep the bond-dimension fixed if one exceeds a predetermined threshold. This provides a general strategy to obtain the matrix product operator representation of the operator $\hat{R}(\hat{\varrho})$.


\section{\label{sec:Appendix3}GHZ-type States}
In this section of the Appendix we discuss GHZ-type states of the form
\begin{equation}
|\psi_{N}(\phi)\rangle = [|0\rangle^{\otimes N/2} |1\rangle^{\otimes N/2} + e^{i\phi} |1\rangle^{\otimes N/2}\rangle |0\rangle^{\otimes N/2}] / \sqrt{2}
\end{equation}
in terms of matrix product states and show how to represent the operator $\hat{R}(\hat{\varrho})$ comprising the POVM elements efficiently as a matrix product operator. Recall that for the GHZ-type state one needs to incorporate two global observables into the set of POVM elements, see equations~\eqref{eqn:GHZPOVM1} and \eqref{eqn:GHZPOVM2} of the main text.

\subsubsection*{The GHZ-type State as a Matrix Product State}
In this subsection we set out to find a matrix product state representation of the GHZ-type state. First, note that
\begin{equation}
|0\rangle^{\otimes N/2} |1\rangle^{\otimes N/2} / \sqrt{2} = \sum_{i_{1},\ldots,i_{N}} P_{1}[i_{1}]\cdots P_{N}[i_{N}] |i_{1},\ldots,i_{N}\rangle
\end{equation}
where
\begin{equation}
\begin{array}{lllll}
P_{1}[1] 	& = &  \frac{1}{\sqrt{2}}	&&  \\
P_{k}[1]	& = & 1				&, \text{ for } k=2,\ldots,N/2 \\
P_{k}[1]	& = & 0				&, \text{ for } k=N/2+1,\ldots,N \\
P_{k}[2]	& = & 0				&, \text{ for } k=1,\ldots,N/2 \\
P_{k}[2]	& = & 1				&, \text{ for } k=N/2+1,\ldots,N. \\
\end{array}
\end{equation}
Similarly, we can find a matrix product state representation with bond-dimension $D=1$ for the second term in the GHZ-type state. Further, adding two matrix product states with bond-dimension $D_{1} = 1 = D_{2}$ results in a matrix product state with bond-dimension $D = 2$. Finally, we have for the GHZ-type state
\begin{equation}
\begin{array}{lllll}
P_{1}[1] 	& = &  \frac{1}{\sqrt{2}} \begin{pmatrix} 1 & 0 \end{pmatrix}	&&  \\
P_{k}[1]	& = & \begin{pmatrix} 1&0\\0&0\end{pmatrix}&, \text{ for } k=2,\ldots,N/2 \\
P_{k}[1]	& = & \begin{pmatrix} 0&0\\0&1\end{pmatrix}&, \text{ for } k=N/2+1,\ldots,N-1 \\
P_{N}[1]	& = & \begin{pmatrix} 0\\1\end{pmatrix}& \\
P_{1}[2] 	& = &  \frac{e^{i\phi}}{\sqrt{2}}\begin{pmatrix} 0&1\end{pmatrix}&&  \\
P_{k}[2]	& = & \begin{pmatrix} 0&0\\0&1\end{pmatrix}&, \text{ for } k=2,\ldots,N/2 \\
P_{k}[2]	& = & \begin{pmatrix} 1&0\\0&0\end{pmatrix}&, \text{ for } k=N/2+1,\ldots,N-1 \\
P_{N}[2]	& = & \begin{pmatrix} 1\\0\end{pmatrix}.& \\
\end{array}
\end{equation}

\subsubsection*{Matrix Product Operator Structure of the $\hat{R}$-Operator}
For the GHZ-type states we suggest to measure on all blocks of $R$ contiguous sites and additionally determine the expectation values of the observables $\hat{\Pi}_{1} = (\id + \hat{X}^{\otimes N})/4$, $\hat{\Pi}_{2} = (\id - \hat{X}^{\otimes N})/4$, $\hat{\Pi}_{3} = (\id + \hat{Y} \otimes \hat{X}^{\otimes N-1})/4$ and $\hat{\Pi}_{4} = (\id - \hat{Y} \otimes \hat{X}^{\otimes N-1})/4$. In this subsection of the Appendix, we show that the operator
\begin{equation}
\hat{R}(\hat{\varrho}) = \frac{1}{M} \sum_{i\in\Delta} \frac{n_{i}}{p_{i}} \hat{\Pi}_{i}  = \frac{1}{M} \sum_{i\in\Delta} \frac{n_{i}}{ \text{tr}[\hat{\Pi}_{i}\hat{\varrho}] } \hat{\Pi}_{i}
\end{equation}
can still be written efficiently in the matrix product operator formalism when incorporating these global observables. Here, $M$ is the total number of measurements, $\Delta$ an index set and $n_{i}$ denotes the number of times outcome $\hat{\Pi}_{i}$ is obtained for all $i\in\Delta$. Now, if $\Delta$ contains all indices corresponding to the local measurements comprised by $\Delta_{1}$ plus the four indices contained in $\Delta_{2}$ labelling the global POVM elements $\hat{\Pi}_{1} = (\id + \hat{X}^{\otimes N})/4$, $\hat{\Pi}_{2} = (\id - \hat{X}^{\otimes N})/4$, $\hat{\Pi}_{3} = (\id + \hat{Y} \otimes \hat{X}^{\otimes N-1})/4$ and $\hat{\Pi}_{4} = (\id - \hat{Y} \otimes \hat{X}^{\otimes N-1})/4$, the operator $\hat{R}$ is given by
\begin{equation}
\hat{R}(\hat{\varrho}) = \hat{R}_{1}(\hat{\varrho}) + \hat{R}_{2}(\hat{\varrho})
\end{equation}
where $\hat{R}_{1}(\hat{\varrho})$ contains the local measurements and $\hat{R}_{2}(\hat{\varrho})$ the four global POVM elements. Now one agrees that $\hat{R}(\hat{\varrho}) $ is indeed a matrix product operator with small bond-dimension. The first term is a matrix product operator with small bond-dimension $D_{1}$ due to its locality (see Appendix \ref{sec:Appendix2}) and the global POVM elements are all matrix product operators with bond-dimensions equal to $1$. Hence, $\hat{R}(\hat{\varrho}) $ is a matrix product operator with bond-dimension equal to $D_{1}+4$. In fact, we can even find a representation of $\hat{R}_{2}(\hat{\varrho})$ with a smaller bond-dimension. We have
\begin{equation}
\hat{R}_{2}(\hat{\varrho}) = \sum_{i=1}^{4} \frac{f_{i}}{p_{i}} \hat{\Pi}_{i} = (\frac{f_{1}}{p_{1}} + \frac{f_{2}}{p_{2}}  + \frac{f_{3}}{p_{3}}  + \frac{f_{4}}{p_{4}} ) \cdot \id^{\otimes N} / 4 + (\frac{f_{1}}{p_{1}} - \frac{f_{2}}{p_{2}}) \cdot \hat{X}^{\otimes N} / 4 + (\frac{f_{3}}{p_{3}} - \frac{f_{4}}{p_{4}}) \cdot \hat{Y}\otimes \hat{X}^{\otimes N-1} / 4. \\
\end{equation}
In this notation, the second and third terms can be combined to a matrix product operator with bond-dimension $1$ such that $\hat{R}_{2}(\hat{\varrho})$ has bond-dimension $D_{2} = 2$. Consequently, $\hat{R}(\hat{\varrho})$ is a matrix product operator with bond-dimension $D_{1}+2$. The matrices of the matrix product operator representation of $c_{1} \cdot \hat{X}^{\otimes N} + c_{2} \cdot \hat{Y}\otimes \hat{X}^{\otimes N-1}$ can be chosen as
\begin{equation}
\begin{array}{lllll}
P_{k}[1] 	& = &  0					&, \text{ for } k=1,\ldots,N \\
P_{1}[2]	& = & \sqrt{2}c_{1} 			&& \\
P_{k}[2]	& = & \sqrt{2} 				&, \text{ for } k=2,\ldots,N \\
P_{1}[3]	& = & \sqrt{2}c_{2} 			&& \\
P_{k}[3] 	& = &   0					&, \text{ for } k=2,\ldots,N \\
P_{k}[4]	& = &  0					&, \text{ for } k=1,\ldots,N \\
\end{array}
\end{equation}
where $1,2,3,4$ correspond to the normalized Pauli basis elements $\hat{P}^{(1)} = \id / \sqrt{2},\hat{P}^{(2)} = \hat{X} / \sqrt{2},\hat{P}^{(3)} = \hat{Y} / \sqrt{2}$ and $\hat{P}^{(4)} = \hat{Z} / \sqrt{2}$. Thus, we have
\begin{equation}
c_{1} \cdot \hat{X}^{\otimes N} + c_{2} \cdot \hat{Y}\otimes \hat{X}^{\otimes N-1} = \sum_{i=2}^{3} P_{1}[i]P_{2}[2]\cdots P_{L}[2] \;\hat{P}_{1}^{(i)}\otimes \hat{P}_{2}^{(2)} \otimes \ldots \otimes \hat{P}_{N}^{(2)}
\end{equation}
such that $\hat{R}_{2}(\hat{\varrho})$ is a sum of two matrix product operators with bond-dimensions equal to $1$.


\begin{thebibliography}{00}
\bibitem{remark1}
We label operators using hats, i.e., $\hat{\varrho}$ for a density matrix, rather than applying this notation to estimates of the underlying system.
\bibitem{remark2}
Due to the multiplication of the matrix product operators and a subsequent compression one has to be careful by implementing the algorithm to maintain the Hermiticity in the current iteration.
\bibitem{remark3}
We use a fourth order Trotter expansion with a step size in the imaginary time evolution of $\delta T=10^{-3}$. We approximate the state as a matrix product operator with bond-dimension $D=100$.
\bibitem{remark4}
The random character of the Hamiltonian suggests that the eigenfunctions of the Hamiltonian are strongly localized and hence the ground state is very well approximated by a matrix product state of low bond-dimension.
\bibitem{remark5}
We approximate the ground states with matrix product states of bond-dimension $D=20$.
\end{thebibliography}
\end{document}